\documentclass[12pt]{article}
\usepackage{epsfig}
\begin{document}
\textwidth 160mm
\textheight 240mm
\topmargin -20mm
\oddsidemargin 0pt
\evensidemargin 0pt
\newcommand{\beq}{\begin{equation}}
\newcommand{\eeq}{\end{equation}}
\newcommand{\be}{\begin{equation}}
\newcommand{\ee}{\end{equation}}
\newcommand{\bea}{\begin{eqnarray}}
\newcommand{\eea}{\end{eqnarray}}
\newcommand{\lll}{\lambda}
\def\Journal#1#2#3#4{{#1} {\bf #2}, #3 (#4)}

% Some useful journal names
\def\NCA{\em Nuovo Cimento}
\def\NIM{\em Nucl. Instrum. Methods}
\def\NIMA{{\em Nucl. Instrum. Methods} A}
\def\NPB{{\em Nucl. Phys.} B}
\def\PLB{{\em Phys. Lett.}  B}
\def\PRL{\em Phys. Rev. Lett.}
\def\PRD{{\em Phys. Rev.} D}
\def\ZPC{{\em Z. Phys.} C}
% Some other macros used in the sample text
\def\st{\scriptstyle}
\def\sst{\scriptscriptstyle}
\def\mco{\multicolumn}
\def\epp{\epsilon^{\prime}}
\def\vep{\varepsilon}
\def\ra{\rightarrow}
\def\ppg{\pi^+\pi^-\gamma}
\def\vp{{\bf p}}
\def\ko{K^0}
\def\kb{\bar{K^0}}
\def\al{\alpha}
\def\ab{\bar{\alpha}}
\def\be{\begin{equation}}
\def\ee{\end{equation}}
\def\bea{\begin{eqnarray}}
\def\eea{\end{eqnarray}}
\def\CPbar{\hbox{{\rm CP}\hskip-1.80em{/}}}%temp replacement due to no font

%\begin{titlepage}

\begin{flushright}
hep-th/0203016 \\
DFAQ-02/TH/01 \\ 
ITEP-TH-10/02\\
LPTHE-02-15  \\
IHES/P/02/12  \\
LPT-Orsay-02/13\\
OUTP-02-05P 
\end{flushright}

\setcounter{footnote}{0}
\setcounter{equation}{0}

\vspace{0.5cm}

\begin{center}
{\Large \bf On the Deconstruction of Time}
\end{center} 

\begin{center}
{\large Z. Berezhiani$^{\,a,b}$,
A. Gorsky$^{\,c,d}$ and I.I. Kogan$^{\,c,e,f,g}$  } 
\end{center} 

\begin{center}
$^a$ {\em Dipartimento di Fisica, Universit\`a di L'Aquila,
I-67010 Coppito AQ, and \\
INFN, Laboratori Nazionali del Gran Sasso, I-67010 Assergi AQ, 
Italy}\\
$^b$ {\em Andronikashvili Institute of Physics, 380077 Tbilisi, 
Georgia} \\
$^c$ {\em Institute of Theoretical and Experimental Physics, \\
B.Cheremushkinskaya 25, 117259 Moscow, Russia }\\
$^d$ {\em LPTHE, Universit\'e Paris VI, 
4 Place Jussieu, Paris, France}\\
$^e$
{\em  Theoretical Physics, Department of Physics, Oxford University\\
1 Keble Road, Oxford, OX1 3NP, UK } \\
$^f$
{\em IHES, 35 route de Chartres, 
91440 Bures-sur-Yvette,  France }\\
$^g$
{Laboratoire de Physique Th\'eorique,  
%\footnote{Unite Mixte de Recherche du CNRS (UMR 8627)},
Universit\'e de Paris XI, \\
91405 Orsay C\'edex, France
}
\end{center}

\vspace{3mm}

\begin{abstract}
In this note we discuss the possibility to get
a time rather than space in the
scenario of (de)construction of new dimension.
\end{abstract}

\vspace{10mm}

1.  Recently it was suggested in refs. 
\cite{Arkani-Hamed:2001ca,Hill:2000mu}
that  a four-dimensional gauge theory with a large gauge symmetry
 behaves in the infrared region in a manner  which is very similar to a
five-dimensional gauge theory  with  a smaller gauge group.
 This leads to  an idea of (De)construction of extra dimensions, i.e
 that  extra dimensions  do  not exist 
at fundamental level  and emerge  dynamically in the infrared limit.
The basic idea of (De)construction is the following
\cite{Arkani-Hamed:2001ca,Hill:2000mu}. 
One starts from a theory with a chain of gauge symmetries  
$G_1 \times G_2 \times ...\times G_N$ 
where all groups $G_i$ are identical, i.e. we have  $N$ copies 
of the same gauge group $G$.  
Matter is represented by a set of  scalar
fields  $\Phi_{i,i+1}$'s each of which is transformed 
as a fundamental representation with respect
to symmetry $G_i$ and anti-fundamental with respect to 
the neighbor $G_{i+1}$.\footnote{
Alternatively, instead of fundamental scalar fields one can 
consider some bilinear fermion condensates 
\cite{Arkani-Hamed:2001ca}, but this is not so important. } 
These scalar fields $\Phi_{i,i+1}$ develop non-zero VEVs 
and hence the total gauge symmetry will be
broken down to a diagonal subgroup G. 
For simplicity, let us consider the case when 
$G = U(1)$, i.e. scalars $\Phi_{i,i+1}$ have charges 
$Q_i=1$ and $Q_{i+1}=-1$ with respect to the neighbor 
groups $U(1)_i$ and $U(1)_{i+1}$. 
The system is described by the Lagrangian 
\be \label{lagr} 
{\cal L}= 
-\frac{1}{4g^2}\sum_{i=1}^N F_{(i)\mu\nu} F_{(i)}^{\mu\nu } -
\sum_{i=1}^{N} D_\mu\Phi_{i,i+1}^\dagger D^\mu\Phi_{i,i+1}
\ee
(the signature $(-,+,+,+)$ is chosen),  
where the covariant derivative is defined as 
$D_{\mu}= \partial_{\mu} + {\rm i} \sum_{i} A_{(i)\mu} T^{(i)}$ 
and  $T^{(i)}$ are the generators of the gauge symmetry with 
respect to the group number $i$. Therefore, 
for a field $\Phi_{i,i+1}$  one gets
\be \label{deriv} 
D_\mu\Phi_{i,i+1}  = \partial_{\mu}\Phi_{i,i+1} + {\rm i}
(A_{(i)\mu} -  A_{(i+1)\mu}) \Phi_{i,i+1} \,. 
\ee
When  order parameter $\Phi_{i,i+1}$ acquires non-zero VEV one has 
\be \label{VEV}
\Phi_{i,i+1}  = v\exp({\rm i}\phi_{i,i+1}/\sqrt{2}gv) \, .
\ee

Neglecting in the IR limit the  kinetic energy
$ \partial_{\mu}v \partial^{\mu}v$,  one can see that 
the scalar contribution to the Lagrangian  equals to
\be \label{same} 
\frac{1}{2g^2}\sum_{i=1}^{N}
\left(\partial_{\mu}\phi_{i,i+1} - \sqrt{2} gv
(A_{(i+1)\mu} -A_{(i)\mu})\right)
\left(\partial^{\mu}\phi_{i,i+1} - \sqrt{2} gv
(A_{(i+1)}^{\mu} -A_{(i)}^{\mu})\right) \, . 
\ee
This term  has the structure of 
a discrete version of $F_{\mu,5}F^{\mu,5}$  where  
phase $\phi_{i,i+1}$ is a phase for
link variable for component $A_5(x,i)$ and
\be \label{F5}
\sqrt{2} gv(A_{(i+1)\mu} -A_{(i)\mu}) \rightarrow
\partial_5 A_{\mu}(x,i) + O(v^{-1})\partial_5^2 A_{\mu}(x,i) \, . 
\ee
The lattice spacing $a$ is related to the condensate $v$ 
through the relation
\be \label{a}
\sqrt{2} gv a =1
\ee
so continuous limit corresponds to large $v$.
In this  model  one gets a ``transverse lattice'' description 
of a full $4+1$ gauge theory where
the  size of extra space  $L$ is proportional to the number 
of independent gauge symmetries in an unbroken phase: 
\be \label{L} 
L = Na = \frac{N}{\sqrt{2} gv }  \, . 
\ee

It was shown recently that the deconstruction survives
nonperturbatively in the supersymmetric case \cite{cek}
and two compact dimensions instead of one can be constructed
along this way if the full lattice of nonperturbative
states in taken into account \cite{ahm}.
Moreover,  recently there have been attempts of introducing 
gravity in this scenario
\cite{Sugamoto:2001uk}.
On other phenomenological applications of this approach
see \cite{phen}.

2. Of course  this construction can be repeated for 
space-time with any dimensions and  starting from 
$d+1$-dimensional {\it space-time} we can get 
$(d+1)+1$ dimensional {\it space-time} in the  infrared limit.  
One can immediately ask the following question: 
is it possible  to start from $d$-dimensional {\it space} 
and get $d+1$ dimensional {\it space-time}?
In other words, can we get time out of nothing using 
(De)constructing ?

The answer on this question appears to be  positive,
but we have to work with the system involving the
apparently tachyonic degrees of freedom.
Actually the different dependence of the
vector and scalar degrees of freedom
on the metric seems to be crucial for our purpose.
Therefore, we explicitly restore the dependence
on the metric in gauge action: 
\be \label{action}
{\cal L}= - \frac{1}{4g^2}\sum_{i=1}^N
g^{mn} g^{kl} F_{(i)mk} F_{(i)nl} -
\sum_{i=1}^{N} g^{mn}D_m\Phi_{i,i+1}^\dagger D_n\Phi_{i,i+1}
\ee
and carefully study the metric which arises after
deconstruction.

Consider first a space-time with the metric 
$g_{nm}={\rm diag}(-++...+)$, in the spirit of the 
example discussed in refs. [1,2]. 
For simplicity, let us consider the case d=3 with
$g_{nm}={\rm diag}(-++)$. 
Lagrangian (\ref{action})  leads to a well-defined action
\be \label{S} 
S = \int d^3 x \sqrt{-g} {\cal{L}} \, . 
\ee
For the choice of metric $g_{nm}={\rm diag}(+--)$, 
we have the wrong sign in front of the scalar part of the action 
(it would be ghost-like), but the gauge part is all right. 
Transition from signature $(-++)$ to the signature $(+--)$ 
is nothing but transformation
\be \label{g-g} 
g_{nm} \rightarrow  - g_{nm}
\ee
and obviously the Lagrangian of vector fields is invariant 
under this transformation, while the kinetic part of 
the scalar Lagrangian is not.
Let us note that the path integral for the theory
with the action (\ref{action}) is defined
with an oscillating exponent
\be \label{path} 
\int D A_m (x) \prod_{i} D\Phi_{i,i+1}(x)
\exp\{ -{\rm i} \int d^3 x \sqrt{-g}{\cal{L}}\} \, . 
\ee 
In the low energy limit this path integral describes a gauge 
(Maxwell or Yang-Mills) theory including a matter 
in a $3+1$ dimensional space-time 
with one extra compact {\it spatial} direction and 
metric $G_{\mu\nu} = {\rm diag}(-+++)$.

Let us turn now now to our main observation and show how  
the time coordinate can be generated if the metric 
different from $g_{nm}={\rm diag}(-++)$ 
is chosen before deconstruction.
The most interesting possibility is to assume that
our metric describes the Euclidean space, 
i.e. all directions have the same signature.
There are formally two possibilities: 
\begin{itemize}
\item 
To take metric $g_{E} = {\rm diag}(+++)$, i.e.
all coordinates are like spatial coordinates  in 
our original $2+1$ space-time. 
(Of course there is no time now - it exists only when we
have quadratic form with indefinite sign.)  
\item 
To take metric $g_{L} = {\rm diag}(---)$, i.e.
in our original $2+1$ `space-time' all coordinates 
are like temporal coordinates, or formally they are 
the spatial ones but the ``kinetic'' terms 
of $\Phi$-fields in (\ref{action}) have the wrong sign.  
\end{itemize}

In the first case one can see that  starting from $g_{E}$, 
after deconstruction one gets extra spatial coordinate 
and hence an Euclidean gauge theory in $d=4$ space: 
\be \label{E} 
G_{E}^{MN}  = \delta^{MN} \, , 
\ee
with the  low-energy action
\be \label{i} 
-{\rm i} S = - {\rm i} 
\frac{1}{4g^2}\int d^3 x dy  \sqrt{G_E}F_{MN}F_{MN}
\ee
where the extra `i' amounts from the factor $\sqrt{-g}$ 
and it cancels with the factor `i' in the path integral 
(\ref{path}), thus yielding an Euclidean field theory with 
a real path integral:  
\be \label{real} 
\int D A_M (x,y) \exp\{ -S[A]\} \, . 
\ee

Let us consider now the second choice, i.e. when we have 
metric $g_{L} = {\rm diag}(---)$. 
It is easy to see that in this case relative signs
of gauge and scalar sectors are different. 
If one considers a lattice regularization for the space one can 
see that scalar part corresponds to antiferromagnetic coupling 
between nearest neighbors, contrary to the
ferromagnetic coupling in the space case: 
the state where neighbor fields are close to each other 
does not not correspond to minimum but to maximum
implying the existence of the unstable mode.
One can easily see that repeating the same steps
we do not change anything in a gauge part 
(which becomes now the magnetic field part of the action) 
but because of the change of the sign in front of scalar part 
we effectively get an electric field contribution.

The naive way to get the electric field contribution 
$F_{0 m}F^{0 m} = - F_{0 m}F_{0 m}$
out of the Goldstone part is just to identify
$\phi_{i,i+1}(x)$ as a scalar potential $A_{0}(x, i)$
and approximate in the expression 
\be \label{Fom}
F_{0 m} (x, i) = 
\partial_{m}\phi_{i,i+1}(x) - \sqrt{2} gv
(A_{(i+1) m}(x) -A_{(i) m}(x))  
\ee
the combination $A_{(i+1) m}(x) -A_{(i) m}(x)$
as a time derivative $\frac{dA_m}{dt}\delta t$ .
In this view the step of the ``time lattice''
is nothing but the inverse value of the condensate, 
$\delta t= 1/gv$. The higher derivative terms are
suppressed by the large value of the condensate.

As a result, we get into the theory in a 
space-time having the Minkowski  metric
\be \label{Mink} 
G_{L}^{\mu\nu}  = {\rm diag}(-,+,+,+) \, . 
\ee
Since the factor $\sqrt{-G}$ should appear
in this case there is no extra `i' when we go from
$3$ to $3+1$ dimensions. Therefore, we derive the correct 
path integral (with complex phase) for the gauge theory 
in Lorentzian space-time. Note also that the deconstruction 
effectively restores the symmetry $G_{MN} \to - G_{MN}$ 
which becomes the effective low energy symmetry. 

One can  be more precise and look more
carefully on the ``mass term'' for the ``W-bosons'',
which can be easily found  from the action. 
In refs. \cite{Arkani-Hamed:2001ca,Hill:2000mu}
this term was immediately identified with KK masses using 
the mode expansion of the gauge field $A(t,\vec{x},x_5)$.
Now, since we are hunting for the time, it is natural
to assume that such term amounts from the mode
expansion of $A_{m}(t,\vec{x})$ in $t$ variable.
As far as the eigenvalues of the mass matrix
have the structure $m_k=gv \sin(k/N)$ at large N
the linear spectrum of frequencies $\omega_k \propto k$
has to be somehow explained. In the KK case it is just the
consequence of the periodic or $Z_2$ orbifold
boundary conditions. However in the case of time 
periodic conditions in the Minkowski space are not 
acceptable. Hence, in our case we could have only a kind 
of orbifold boundary conditions on the ``boundary of the Universe''
or free boundary conditions and infinite N.  
In principle one can consider finite $N$  and this case 
will correspond to the ``Universe'' which originates at 
some moment and which existence will be terminated
at some later moment.  
Perhaps this approach could be also useful to discuss
the periodic time, for
example the case of $AdS$ space-time.

Since the time direction emerges dynamically we have to 
examine the Gauss law selecting the gauge invariant states 
at the quantum level.
Let us compare how  Gauss law and gauge invariance are 
realized in the KK and ``time'' cases.
In first case we have
\bea
\partial_m E_{(i)m} & = & \rho_{i},   \nonumber \\
-\rho_i & = & \partial_{0}\phi_{i,i+1}(x) - \sqrt{2} gv
(A_{(i+1) 0}(x) -A_{(i) 0}(x))   \nonumber \\
 & - & \partial_{0}\phi_{i-1,i}(x) - \sqrt{2} gv
(A_{(i) 0}(x) -A_{(i-1) 0}(x))  \, . 
\eea
One can easily see that in a continuum limit the 
density $\rho_{i}$ becomes nothing but $-\partial_5 E_5$. 
As a result we get a five dimensional Gauss law
\be \label{Gauss} 
\partial_m E_{m} + \partial_5 E_5 = 0 \, . 
\ee

The totally different story appears when we want 
to get time out of deconstruction. 
In this case we do not have  electric field to start with. 
One can see that the  Gauss law  $\partial_m E_{m} =0$ 
which is supposed to be valid at any
time moment $i$ can be written as
\be \label{Gauss-t} 
\partial_m F_{0 m} (x, i) = 
\partial^2\phi_{i,i+1}(x) - \sqrt{2} gv
(\partial_{m} A_{m}(x, i+1) - \partial_{m}A_{ m}(x, i)) =0
\ee
One can choose all $\phi_{i,i+1}(x) =0 $ 
(this is $A_0 =0 $ gauge). 
The Gauss law in this case reads as
\be 
\partial_{m} A_{m}(x, i+1)  = \partial_{m}A_{m}(x, i).
\ee
and corresponds to the time independent gauge fixing.

Let us also comment on the possible relation between the 
quasiclassical nonperturbative configurations.
It was argued in KK supersymmetric case that nonperturbative
configurations are mapped to each 
other under deconstruction \cite{cek}. 
In the `time' case we would like to get, for instance, 
instanton in the deconstructed theory. 
It can be obtained indeed, considering the infinite arrow of 
monopoles in d=3 theory along the "time" direction.

3. Now let us briefly mention what are the physical consequences 
of the picture proposed.
First note that the perception of time as a chain of 
the ordered events with no return to the past is just 
the fact that we are measuring observables in the $i$-th 
sector only once. 
After we  measured it we have to measure the next one -- 
and never can return back -- because the wave function 
in that sector is already defined. 
Second, it is natural to ask if a kind of appearance 
(or disappearance) of the time dimension in the IR(UV) limits 
similar to what happens in the KK case is possible. 
In the KK case the condensate can be destroyed at large energies, 
yielding the effective disappearance of the fifth dimension. 
One can relate this with the uncertainty relation 
$\Delta x_5 \Delta p_5 \geq 1$ -- when we try to localize 
the position in space we bring such an uncertainty in the 
momentum that causes the destruction of the condensate. 
If for example we shall destroy the condensate on the link
$(i, i+1)$ it will cause the creation of two disconnected 
worlds.  
In the ``time'' case one could also imagine the ``dynamical'' 
disappearance of the condensate which would look like that  
``time disappears''. 
Again, if we try to destroy a condensate just on one link, 
it will cause also emergence of two disconnected worlds - 
but now they will be disconnected in time. 
Now we have  the uncertainty relation
$\Delta E \Delta t \geq 1$, and because we have 
the time quanta $\Delta t \sim v^{-1}$, 
one can not have $\Delta E \leq v$.

Our perception of time is a causally ordered sequence of the 
processes of measurement each of which can happen only once. 
If time emerges in the way as we have just described, this 
would mean that the observer can make a measurement for all 
gauge fields - but only once.
In some sense  the full evolution is just one complete measurement.
This sounds quite natural indeed.  
When one is asking question about what happens when 
we repeat measurement it is based on the assumption that we
can measure something again later. 
But later means later in time - and if time itself emerges 
dynamically this question simply can not be asked. 
There is only {\it one} measurement for each sector - 
which means that you can not return back in the  past and 
reobserve the things.

Among other interesting points to be  questioned is the
deconstruction of d=2 YM theory. Since in d=2 
the YM theory is topological, only zero modes on the cylinder
are relevant and all higher KK modes can be cut off safely. 
The theory which amounts to d=2 YM theory after deconstruction 
can be presented as $N$ copies of quantum mechanics
where $N$ defines the radius of the cylinder.
However, since only zero mode works, only one copy
of quantum mechanics is relevant. 
Now turn to the question concerning the signature of d=2 theory. 
Before deconstruction, the issue of signature
in quantum mechanics is subtle since we have to deal with 
``world-line'' (time) and ``target'' (coordinate) simultaneously. 
It can be well defined only for the relativistic particle since
the length of the world-line is defined with some metric.

On the other hand d=2 YM theory 
at large N is equivalent to c=1 string theory,    
that is what we are talking about is the deconstruction of c=1 
string theory from the set of copies of quantum mechanics. 
Moreover, from the viewpoint of c=1 string we are 
deconstructing the {\it target manifold} since d=2 YM theory 
is defined  on the target from the stringy point of view. 
The way how the second dimension emerges in c=1 string is known --  
it is the Liouville mode while the c=1 string theory
can be defined via matrix  quantum mechanics indeed.
Hence the issue of the resulting metric in d=2
theory is related with the sign in front of the
Liouville contribution to the action. Usually
it is assumed that Liouville field plays
the role of time.
One more potential question concerns
the deconstruction of (0+1) theory (quantum mechanics)
from the copies of (0+0) (matrix model).
This has something to do with the M(atrix) model
deconstruction of D0 brane from D-instantons.

It is known how the deconstruction procedure can be 
formulated in terms of branes.  For instance, to get the quiver models
 one could take 
the set  of D3 branes on orbifolds in SUSY case \cite{dm}
then $W$ bosons are represented 
by the strings connecting the pairs of D3 branes.
When this  paper was almost completed 
 the  preprint  \cite{gutstrom} appeared where 
new branes localized in time direction have been found.
These branes are very natural objects to start with to get 
the new time like-coordinates in terms of the brane array leading to
 the group products. 
Since fundamental strings can end on them the spectrum of 
"masses" could be reproduced in a way similar to the KK case.

4. In conclusion let us make our main statement again. 
Starting from the action (\ref{action}), 
it is possible to get a quantum field theory 
in a space-time where either extra spatial coordinate or 
extra time emerges via deconstruction. 
Alternatively one can get statistical field theory.

\vspace{0.5cm}

{\bf Acknowledgments} 
\vspace{2mm}

The work of Z.B. is partially supported by the MURST research 
grant "Astroparticle Physics", the work A.G. is supported 
in part by grants INTAS-00-00334  and CRDF-RP2-2247,
and that of I.I.K. is supported in part by PPARC
rolling grant PPA/G/O/1998/00567 and EC TMR grant
HPRN-CT-2000-00152.

\end{document}